\documentclass[aps,prd,showpacs,nofootinbib,10pt,superscriptaddress,twocolumn,amsmath,amssymb,floatfix,showkeys]{revtex4-1}
\usepackage{color,amsmath,amssymb,graphicx,latexsym,lineno}
\usepackage{threeparttable,multirow,txfonts,booktabs}
\usepackage{accents,slashed,ulem,subfig}
\usepackage[colorlinks,linkcolor=blue,anchorcolor=blue,citecolor=blue,urlcolor=blue]{hyperref}
 \usepackage{float}
 \usepackage{flushend}
 \usepackage{balance}

\newlength{\dhatheight}
\newcommand{\doublehat}[1]{%
    \settoheight{\dhatheight}{\ensuremath{\hat{#1}}}%
    \addtolength{\dhatheight}{-0.15ex}%
    \hat{\vphantom{\rule{1pt}{\dhatheight}}%
    \smash{\hat{#1}}}}
\graphicspath{{figs/}}

\begin{document}
\title{Constraints on Axion-like Particles from the Observation of Galactic Sources by LHAASO}
\author{Jun Li}
\affiliation{Key Laboratory of Dark Matter and Space Astronomy, Purple Mountain Observatory, Chinese Academy of Sciences, 210033 Nanjing, Jiangsu, China}
\affiliation{School of Astronomy and Space Science, University of Science and Technology of China, 230026 Hefei, Anhui, China}

\author{Xiao-Jun Bi}
\affiliation{Key Laboratory of Particle Astrophysics, Institute of High Energy Physics,
Chinese Academy of Sciences, Beijing 100049, China}
\affiliation{School of Physical Sciences, University of Chinese Academy of Sciences, Beijing, China}
\author{Lin-Qing Gao}
\affiliation{School of Nuclear Science and Technology, University of South China, Hengyang 421001, China}

\author{Xiaoyuan Huang}
\affiliation{Key Laboratory of Dark Matter and Space Astronomy, Purple Mountain Observatory, Chinese Academy of Sciences, 210033 Nanjing, Jiangsu, China}
\affiliation{School of Astronomy and Space Science, University of Science and Technology of China, 230026 Hefei, Anhui, China}

\author{Run-Min Yao}
\affiliation{Key Laboratory of Particle Astrophysics, Institute of High Energy Physics,
Chinese Academy of Sciences, Beijing 100049, China}
\affiliation{School of Physical Sciences, University of Chinese Academy of Sciences, Beijing, China}

\author{Peng-Fei Yin}
\affiliation{Key Laboratory of Particle Astrophysics, Institute of High Energy Physics,
Chinese Academy of Sciences, Beijing 100049, China}
\email[E-mail: ]{yinpf@ihep.ac.cn}

\date{\today}

\begin{abstract}

High-energy photons may oscillate with axion-like particles (ALPs) when they propagate through the Milky Way's magnetic field, resulting in an alteration in the observed photon energy spectrum. The ultra-high energy gamma-ray spectra, measured by the Large High Altitude Air Shower Observatory (LHAASO) up to $\mathcal{O}(1)~\mathrm{PeV}$, provide a promising opportunity to investigate the ALP-photon oscillation effect. In this study, we utilize the gamma-ray spectra of four Galactic sources measured by LHAASO, including the Crab Nebula, LHAASO J2226+6057, LHAASO J1908+0621, and LHAASO J1825-1326, to explore this effect. We employ the $\rm CL_s$ method to set constraints on the ALP parameters. 
Combing the observations of the four sources, our analysis reveals that the ALP-photon coupling $g_{a\gamma}$ is constrained to be smaller than
$1.4\times10^{-10}$ ${\rm GeV}^{-1}$ for the ALP mass of $\sim 4\times10^{-7} ~\mathrm{eV}$ at the 95\% C.L.
By combing the observations of the Crab Nebula  from LHAASO and other experiments, we find that the ALP-photon coupling could be set to be about $7.2\times10^{-11}$ ${\rm GeV}^{-1}$ 
for the ALP mass $\sim 4 \times10^{-7}~\mathrm{eV}$ , which is in close proximity to the CAST constraint.

\end{abstract}

\maketitle

\section{introduction}

Axion-Like Particles (ALPs) \cite{choi2021recent, irastorza2018new,de2008axion, graham2015experimental}, a class of pseudo-scalar bosons, arise as a consequence of symmetry breaking in many extensions of the Standard Model. ALPs possess a broader parameter space and a rich phenomenology that is yet to be fully explored, compared to the quantum chromodynamics axions addressing the strong CP problem \cite{Peccei:1977ur,Peccei:2006as,Weinberg:1977ma, Wilczek:1977pj}. 
The effective coupling between the ALP and photons can lead to ALP-photon oscillation in an external magnetic field. This phenomenon has drawn significant attention in astrophysics due to the ubiquitous astrophysical magnetic fields \cite{Raffelt:1987im,DeAngelis:2007dqd, Hooper:2007bq, Simet:2007sa, Mirizzi:2007hr, Mirizzi:2009aj, Belikov:2010ma, DeAngelis:2011id, Horns:2012kw, HESS:2013udx, Meyer:2013pny, Tavecchio:2014yoa, Meyer:2014epa, Meyer:2014gta, Fermi-LAT:2016nkz, Meyer:2016wrm, Berenji:2016jji, Galanti:2018upl, Galanti:2018myb, Zhang:2018wpc, Liang:2018mqm, 
Xia:2018xbt,
bi2021axion, Guo:2020kiq, Li:2020pcn, Li:2021gxs, Cheng:2020bhr, Liang:2020roo, Gao:2023dvn,Gao:2023und, Majumdar:2018sbv,Xia:2019yud,Liang:2018mqm,eckner2022first, galanti2023alp}.

The oscillation between ALPs and photons has the potential to induce irregularities in the gamma-ray spectrum. 
Detecting this phenomenon is typically more achievable at lower energies because detectors exhibit superior energy resolution for low-energy photons compared to high-energy photons. Nevertheless, the impact of ALP-photon oscillation on the astrophysical gamma-ray spectrum could also become pronounced at high energies. Interactions involving high-energy photons and low-energy background photons, such as those originating from the interstellar radiation field (ISRF) \cite{mathis1983interstellar, moskalenko2006attenuation, porter2006inverse}, cosmic microwave background (CMB) \cite{vavryvcuk2018universe}, and extragalactic background light   \cite{finke2010modeling, cooray2016extragalactic, dwek2013extragalactic, bernstein2007optical}, lead to the absorption of high-energy photons, thereby attenuating the observed gamma-ray spectra at high energies. The conversion of photons to ALPs may mitigate this absorption effect, as a result of the interaction between ALPs and high-energy photons.  Consequently, the presence of ALPs would lead to a modification of the expected gamma-ray spectrum at high energies within the standard astrophysical framework.

In recent years, significant progress in high-energy gamma-ray observation experiments has led to remarkable measurements of high-energy gamma-ray spectra
\cite{cao2021ultrahigh,lhaaso2021peta,lhaaso2021performance,cao2023first,breuhaus2022pulsar,abeysekara20172hwc,malone2019first,abeysekara2020multiple,albert2021spectrum,cao2021discovery}.
Notably, the Large High Altitude Air Shower Observatory (LHAASO) \cite{ma2022lhaaso} has contributed significantly to this field. In 2021, the LHAASO collaboration reported the detection of ultra-high energy gamma-ray from the Crab Nebula \cite{lhaaso2021peta}. This measurement encompassed results from two detectors, namely the Water Cherenkov Detector Array (WCDA) and the Kilometer Square Array (KM2A), 
offering a precise gamma-ray spectrum of the Crab Nebula that spans more than three energy orders, from $500~\mathrm{GeV}$ to $1.1~\mathrm{PeV}$. In the same year, the LHAASO collaboration reported the detection of over 530 photons with energies above $100~\mathrm{TeV}$ and up to $1.4~\mathrm{PeV}$ from 12 ultra-high-energy gamma-ray sources \cite{cao2021ultrahigh}. The energy spectra of four Galactic sources, namely the Crab Nebula, LHAASO J2226+6057,
LHAASO J1908+0621, and LHAASO J1825-1326, have been provided in the same report \cite{cao2021ultrahigh}. These measurements of high-energy gamma-ray spectra present a promising opportunity to investigate the ALP-photon oscillation effect.

In this study, we utilize the LHAASO observations from the Crab Nebula, LHAASO J2226+6057, LHAASO J1908+0621, and LHAASO J1825-1326, to impose constraints on the ALP parameters. We consider the absorption effect induced by the CMB and ISRF photons on ultra-high-energy photons within the Milky Way. While several studies 
\cite{bi2021axion, Xia:2018xbt, Xia:2019yud, Majumdar:2018sbv} 
have derived constraints on the ALP parameters from high energy gamma-ray observations of Galactic sources, we emphasize that the photons from the four Galactic sources considered in this study are more energetic. Furthermore, we employ the $\rm CL_s$ method \cite{Junk:1999kv,Read:2002hq_cls} to establish robust constraints on the ALP parameters as \cite{Gao:2023dvn,Gao:2023und}. To further enhance the constraints, we also conduct a combined analysis, incorporating the Crab observations from many other experiments, including HAWC \cite{abeysekara2019measurement}, ${\rm AS{\gamma}}$ \cite{amenomori2019first}, HEGRA \cite{aharonian2004Crab}, MAGIC \cite{aleksic2016major}, HESS \cite{aharonian2006observations}, and VERITAS \cite{meagher2015six}.  

This paper is organized as follows. In Section \ref{ALPPhotonOscillation}, we introduce the ALP-photon oscillation effect and the absorption of high-energy photons in the Milky Way. In Section \ref{Method}, we describe the process of fitting the gamma-ray spectra and the $\rm CL_s$ method. In Section \ref{Results}, we present the constraints on the ALP parameters from the LHAASO observations of four Galactic sources and the combined constraint from the observations of the Crab Nebula by multiple experiments. 
Finally, we conclude in Section \ref{conclusion}.

\section{ALP-photon oscillation}
\label{ALPPhotonOscillation}

In this section, we provide a brief introduction to the ALP-photon oscillation effect in Galactic magnetic fields.
The interaction between the ALP and photons can be described by the Lagrangian term
\begin{equation}
\mathcal{L}_{a\gamma} = -\frac{1}{4} g_{a\gamma} a F_{\mu \nu} \Tilde{F}^{\mu \nu} = g_{a\gamma} a \vec{E}\cdot \vec{B},
\end{equation}
where $g_{a \gamma}$ denotes the coupling coefficient between the ALP and photons, $a$ denotes the ALP field, $F$ denotes the electromagnetic field strength tensor, $\Tilde{F}$ denotes its dual tensor, $\vec{E}$ denotes the photon electric field, and $\vec{B}$ denotes the magnetic field.
The propagation equation for a monochromatic ALP/photon beam can be written as \cite{Raffelt:1987im}
\begin{equation}\label{equ:EOM}
    \left(i \frac{\rm d}{{\rm d} z} + E + \mathcal{M}^{'} \right) \Psi(z) = 0,
\end{equation}
where $z$ denotes the distance along the propagation direction $\hat{z}$, $E$ denotes the energy of ALP/photon, and ${\Psi}\equiv {(A_\perp,A_\parallel,a)}^T$ with $A_\perp$ and $A_\parallel$ representing the photon polarization amplitudes perpendicular and parallel to the transverse component of the external magnetic field $B_t$, respectively. The matrix $\mathcal{M}^{'}$ encompasses the ALP-photon oscillation effect and the absorption effects of high-energy photons, and can be written as
\begin{equation}\label{equ:M}
    \mathcal{M}^{'} = \mathcal{M}+i\left(\begin{array}{lll}
	   \Gamma_{\rm BG} &  & \\
     & \Gamma_{\rm BG} & \\
     &  & 0\\
	 \end{array} \right).
\end{equation}

The mixing matrix  $\mathcal{M}$ includes the interaction between photons and ALPs and environment effects, and is expressed as
\begin{equation}
    \mathcal{M} = \begin{bmatrix}
\Delta_{\perp} & 0 & 0 \\
0 & \Delta_{\parallel} & \Delta_{a\gamma} \\
0 & \Delta_{a\gamma}  & \Delta_{a} 
\end{bmatrix},
\end{equation}
with $\Delta_{\perp} = \Delta_{\rm pl} +2\Delta_{\rm QED}$,  $\Delta_{\parallel} = \Delta_{\rm pl} + 7/2\Delta_{\rm QED}$, $\Delta_{a}=-m_a^2/(2E)$, and $\Delta_{a\gamma}=g_{a\gamma}B_t/2$. Here, $m_a$ is the mass of the ALP. The term $\Delta_{\rm pl} = -\omega_{\rm pl}^{2}/(2E)$ describes the effective mass of photons in plasma with the typical frequency $\omega_{\rm pl}$. $\Delta_{\rm QED}=\alpha E/(45\pi)(B_{t}/B_{\rm cr})^2$ is the QED vacuum polarization term, where $\alpha$ is the fine structure constant, $m_e$ is the electron mass, and $B_{\rm cr}\equiv m_e^2/|e|$ is the critical magnetic field. The off-diagonal element $\Delta_{a\gamma}=g_{a\gamma}B_t/2$ describes the ALP-photon mixing effect. 

When ultra-high-energy photons propagate in the Galactic magnetic field, the absorption effect induced by CMB and ISRF photons through the pair production process $\gamma + \gamma_\text{bkg} \to e^+ + e^-$ \cite{esmaili2015gamma,moskalenko2006attenuation,porter2018galactic} can not be neglected. 
$\Gamma_{\rm BG}$ representing these effects is calculated as 
\begin{equation}\label{equ:lambda}
\Gamma_{\rm BG}=\frac{1}{2}\int dE_{\rm BG} \frac{dn_{\rm BG}}{dE_{\rm BG}} \hat{\sigma}, 
\end{equation}
where $E_{\rm BG}$ and $n_{\rm BG}$  are the energy and number density of background radiation fields, respectively. In this analysis, we utilize the ISRF model presented in Ref. \cite{vernetto2016absorption}.  The term $\hat{\sigma}$ is given by
\begin{equation}\label{equ:sigma}
\hat{\sigma}=\int_{0}^{2}dx\frac{x}{2}\sigma_{\gamma\gamma}, 
\end{equation}
where $x\equiv 1-{\rm cos}{\theta_{\gamma\gamma}}$, and $\theta_{\gamma\gamma}$ is the angle between incident photons. The cross section of the pair production $\sigma_{\gamma\gamma}$ is given by
\begin{equation}\label{equ:sigmaGamma}
\sigma_{\gamma\gamma}=\frac{3}{16}\sigma_{T}(1-\beta^2) \left[(3-\beta^4)\ln{\frac{1+\beta}{1-\beta}}-2\beta(2-\beta^2)\right], 
\end{equation}
where $\beta \equiv(1-4m_e^2/s)^{1/2}$,  $\sigma_{T}$ is the Thomson cross section \cite{canuto1971thomson,borner1979classical,herold1979compton}, and $s=2xEE_{\rm BG}$ is the center-of-mass energy. For the photons from the Crab Nebula with energies of $1~\mathrm{PeV}$, this absorption would result in a loss of $\sim$ 19$\%$ of the photon flux.

The generalized density matrix $ \rho \equiv \Psi \otimes \Psi^\dagger$ can be used to describe the polarized states of the ALP-photon system. This matrix $\rho$
obeys the Von Neumann-like equation \cite{DeAngelis:2007dqd,Mirizzi:2009aj}
\begin{equation}\label{equ:von Neumannn-like}
    i\frac{{\rm d}\rho}{{\rm d}z} = [\rho, \mathcal{M}^{'}].
\end{equation}
The solution to Eq.~\ref{equ:von Neumannn-like} in a homogeneous magnetic field can be expressed as $\rho(z) = \mathcal{T}(z) \rho(0) \mathcal{T}^\dagger(z)$, where the transfer function $\mathcal{T}(z)$ is obtained from the solution of Eq.~\ref{equ:EOM}. 
High energy photons emitted from Galactic sources undergo the magnetic field of
the Milky Way before reaching the Earth. The entire path can be divided into many pieces, with the magnetic field in each domain considered to be homogeneous. The total transfer matrix is given by
\begin{equation}\label{equ:T}
\mathcal{T}(z) = \prod \limits_i^n \mathcal{T}_i (\Delta z_i), 
\end{equation}
where 
$ \mathcal{T}_i (\Delta z_i)$ 
represents the transfer matrix in the $i$-th piece. 

In this study, we focus on the Galactic sources and investigate the ALP-photon oscillation effect in the Galactic field. The Galactic magnetic field consists of a regular component and a turbulent component. As the later component is small and can be safely ignored, we only consider the regular component. The Galactic magnetic model utilized in this study is the Jansson \& Farrar model \cite{jansson2012new}. Additionally, we take the NE2001 model \cite{Cordes:2002wz} for the Galactic electron distribution.

The survival probability of the photon can be expressed as \cite{Raffelt:1987im,Mirizzi:2007hr}
\begin{equation}\label{equ:P_ga}
P_{\gamma\gamma} = \mathrm{Tr}\left((\rho_{11}+\rho_{22})\mathcal{T}(z) \rho(0) \mathcal{T}^{\dagger}(z)\right), 
\end{equation}
where $\rho(0) = \mathrm{diag}(1/2, 1/2, 0)$, $\rho_{11}=\mathrm{diag} (1, 0, 0)$, and $\rho_{22}=\mathrm{diag} (0, 1, 0)$ for unpolarized photons. Considering the ALP-photon oscillation and the absorption effects described above, the observed photon energy spectrum is given by 
\begin{equation}
\label{equ:df/dE} \frac{{\rm d}N}{{\rm d}E} = P_{\gamma\gamma}\left.\frac{{\rm d}N}{{\rm d}E}\right\vert_{\rm int}, 
\end{equation} 
where $\left.\frac{{\rm d}N}{{\rm d}E}\right\vert_{\rm int}$ is the intrinsic spectrum of the source. For the sources considered in this study, we set their intrinsic spectra to be a log-parabolic function given by $F_0(E/E_0)^{-\Gamma - b\mathrm{log} (E/E_0)}$, where $F_0$, $\Gamma$, and $b$ are free parameters, and $E_0$ is set to be $10 ~\mathrm{TeV}$. This choice is based on the results in Ref. \cite{cao2021ultrahigh} and \cite{lhaaso2021peta}, where the log-parabolic spectrum provides better fitting results.

\section{Method}
\label{Method}
In this section, we briefly introduce the analysis method used to set constraints on the ALPs parameters. The best-fit spectrum can be obtained by minimizing the $\chi^2$ function

\begin{equation}
    \chi^2 =  \sum\limits _{j} \chi^2_{j},
\end{equation}
where $\chi^2_{j}$ denotes the $\chi^2$ function of the $j$-th source. $\chi^2_{j}$ is given by
\begin{equation}
    \chi^2_{j}= \sum\limits _{i}\frac{(\widetilde{\Phi}_i-\Phi_{i})^2}{\delta {\Phi_{i}}^2},
\end{equation}
 where $\widetilde{\Phi}_i$, $\Phi_i$, and $\delta \Phi_i$ represent the predicted value, observed value, and experimental uncertainty of the photon flux in the $i$-th energy bin, respectively.

For given $m_a$ and $g_{a\gamma}$, we define the test statistic (TS) as 
\begin{equation}\label{equ:TS_CLs}
    {\rm TS}(m_a, g_{a\gamma}) = \chi^2_{\rm ALP} (\doublehat{F_0}, \doublehat{\Gamma}, \doublehat{b}; m_a, g_{a\gamma})-\chi^2_{\rm Null}(\hat{F_0}, \hat{\Gamma}, \hat{b}),
\end{equation}
where $\chi^2_{\rm Null}$ represents the best-fit $\chi^2$ value under the null hypothesis without the ALP-photon oscillation effect,  $\chi^2_{\rm ALP}$ represents the best-fit $\chi^2$ value under the alternative hypothesis including the ALP-photon oscillation effect with the given two parameters $m_a$ and $g_{a\gamma}$, and $(\hat{F_0}, \hat{\Gamma}, \hat{b})$ and $(\doublehat{F_0}, \doublehat{\Gamma}, \doublehat{b})$ denote the best-fit values of the parameters of the intrinsic spectrum under the null and alternative hypotheses, respectively.

Due to the non-linear impact of ALPs on the photon spectrum, the application of Wilks' theorem \cite{wilk} is unsuitable here, as discussed in \cite{Fermi-LAT:2016nkz}.  As a result, the TS distribution can not be adequately described by a $\chi^2$ distribution, necessitating Monte Carlo simulations to obtain a realistic TS distribution.
In this study, we employ the ${\rm CL_s}$ method \cite{Junk:1999kv,Read:2002hq_cls,Lista:2016chp} to establish constraints on the ALP parameters. The constraints are derived following the same procedure described in Ref.~\cite{Gao:2023dvn,Gao:2023und}. Here, we only offer a concise introduction to this method. 

We explore the $(m_a, g_{a\gamma})$ parameter space, and assess the exclusion of each parameter point using the ${\rm CL_s}$ method in accordance with observations. For each parameter point, we generate two mock data sets, denoted as $\rm \{d\}_{s+b}$ and $\rm \{d\}_{b}$, based on the expected spectra with and without ALPs, respectively. Both mock data sets $\rm \{d\}_{s+b}$ and $\rm \{d\}_{b}$ consist of 1000 samples. For a specific mock data sample, in each energy bin, the photon flux is randomly generated from a Gaussian distribution, where the mean value and deviation are set to be the expected flux and experimental uncertainty, respectively. Utilizing $\rm \{d\}_{s+b}$ and $\rm \{d\}_{b}$, we obtain two TS distributions $\rm \{TS\}_{b}$ and $\rm \{TS\}_{s+b}$ using Eq. \ref{equ:TS_CLs}. Given the TS value ${\rm TS_{obs}}$ obtained from the actual observed data, the $\rm CL_s$ value is defined as 
\begin{equation}
\label{CLs}
   \rm CL_s=\frac{CL_{s+b}}{CL_{b}} ,
\end{equation}
where $\rm CL_{s+b}$ and $\rm CL_b$ represent the probabilities of finding a TS value larger than  ${\rm TS_{obs}}$ according to the distributions $\rm \{TS\}_{s+b}$ and $\rm \{TS\}_{b}$, respectively. If $\rm CL_s$ is less than 0.05, this parameter point is considered to be excluded at a 95\% confidence level (C.L.). 

\begin{figure*}[htbp]
  \centering
  
\includegraphics[width=0.42\textwidth]{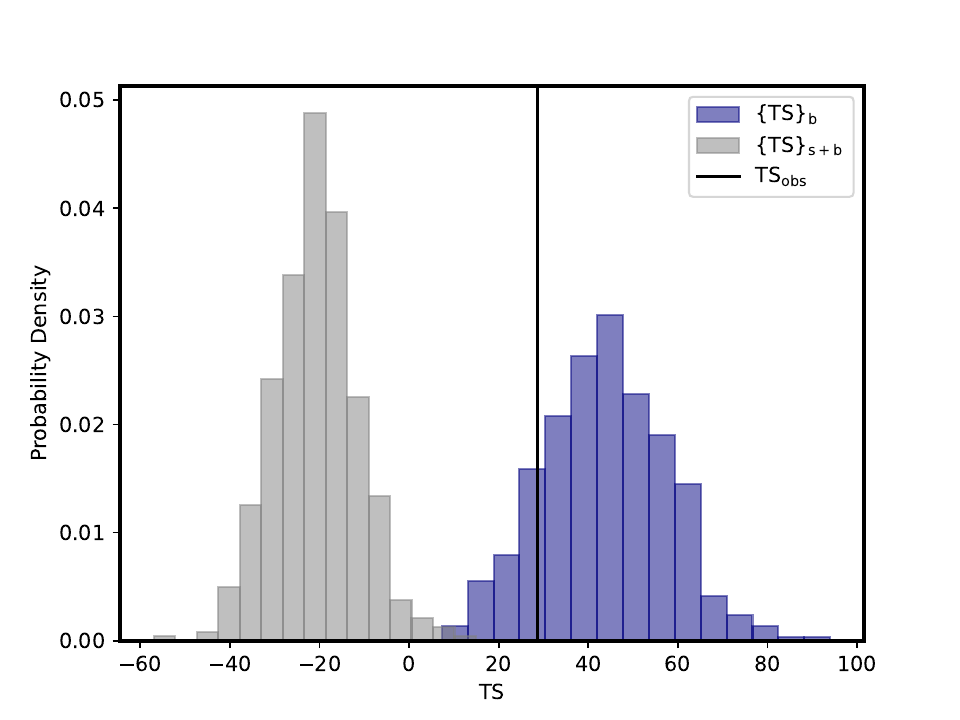}
\includegraphics[width=0.42\textwidth]{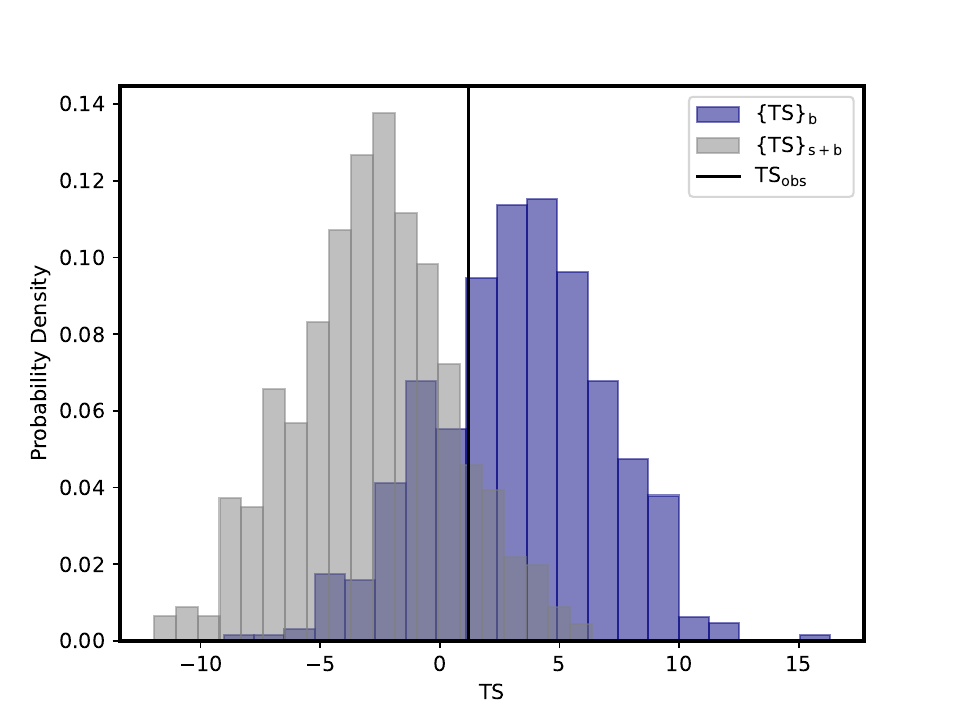}
  \caption{The TS distributions $\rm \{TS\}_{b}$ and $\rm \{TS\}_{s+b}$ for two parameter points $(m_a, g_{a\gamma})= (8\times 10^{-7} \rm eV, 3\times 10^{-10} \rm GeV^{-1})$ and $(3\times10^{-7} \rm eV, 10^{-10} \rm GeV^{-1})$ are shown in the left and right panels, respectively. The vertical black solid lines represent $\rm TS_{obs}$. }
   \label{TSdistri}
\end{figure*}

In FIG. \ref{TSdistri}, we show the TS distributions $\rm \{TS\}_{s+b}$ and $\rm \{TS\}_{b}$ for two specific parameter points $(m_a, g_{a\gamma})= (8\times 10^{-7} \rm eV, 3\times 10^{-10} \rm GeV^{-1})$ and $(3\times10^{-7} \rm eV, 10^{-10} \rm GeV^{-1})$ as examples. 
The corresponding $\rm CL_s$ values for these two parameter points are 0.0 and 0.16, respectively. These result indicate that the first parameter point can be excluded at the 95$\%$, while the second parameter point is still allowed. 

\section{Results}
\label{Results}

\begin{figure*}[htbp]
  \centering
  
\includegraphics[width=0.42\textwidth]{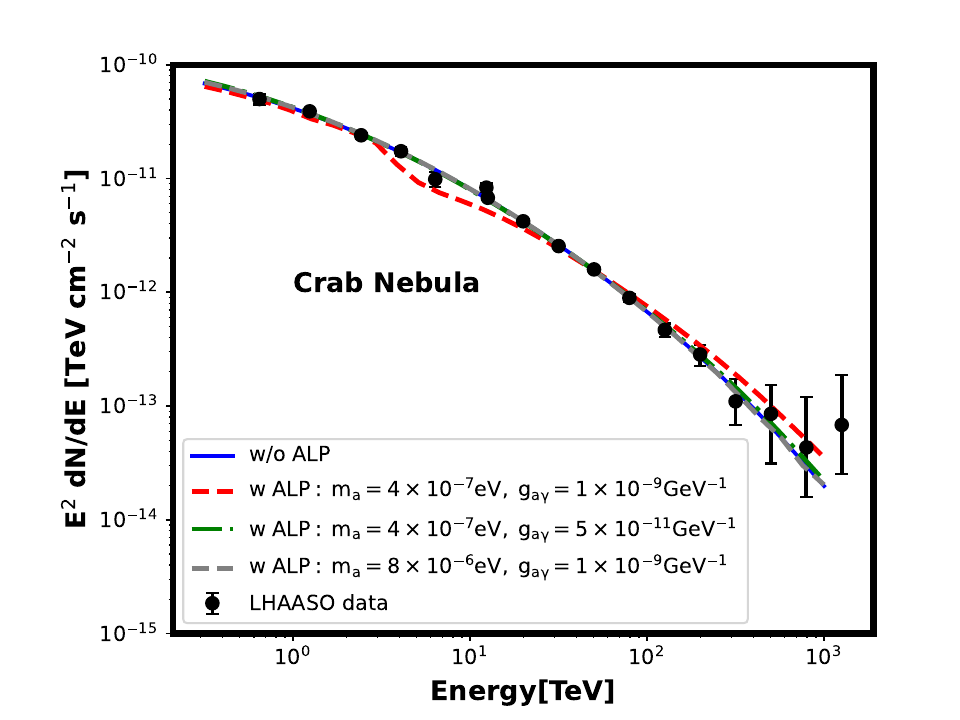}
\includegraphics[width=0.42\textwidth]{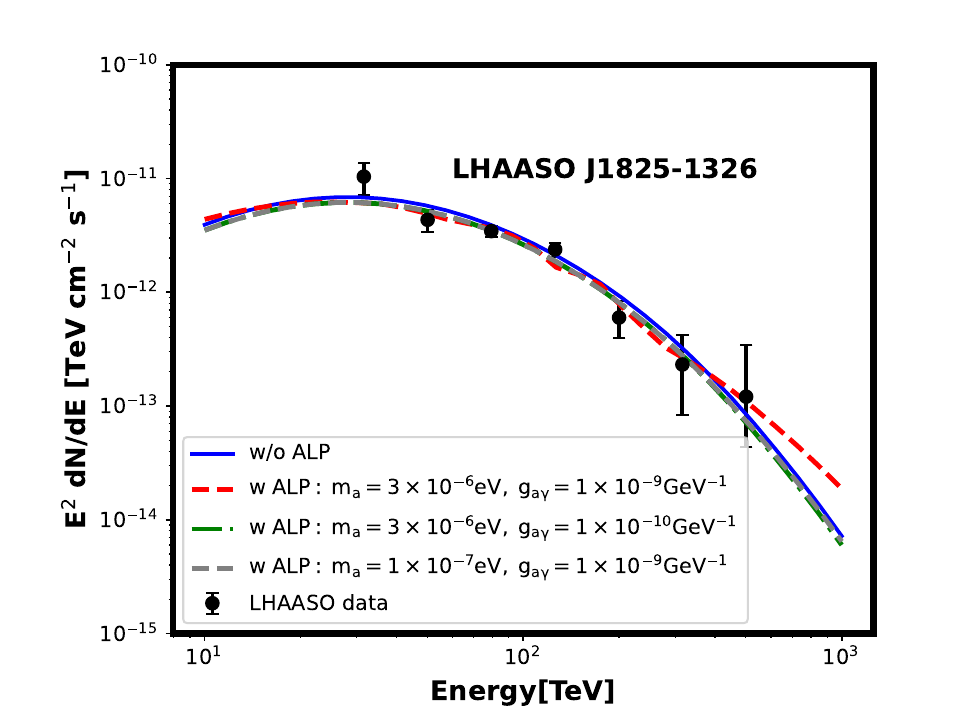}
\includegraphics[width=0.42\textwidth]{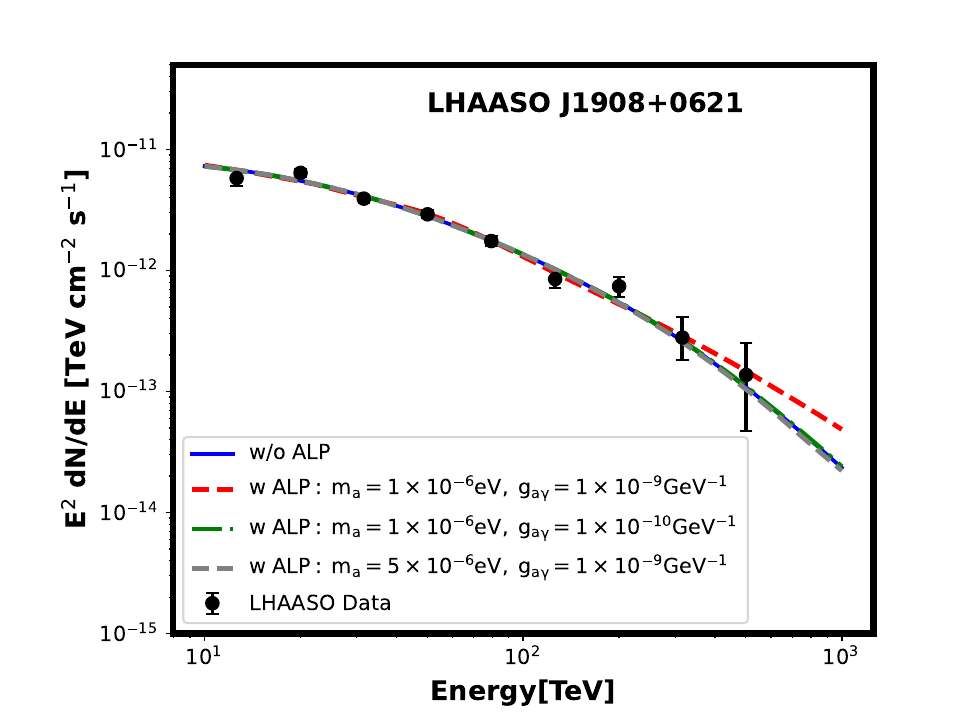}
\includegraphics[width=0.42\textwidth]{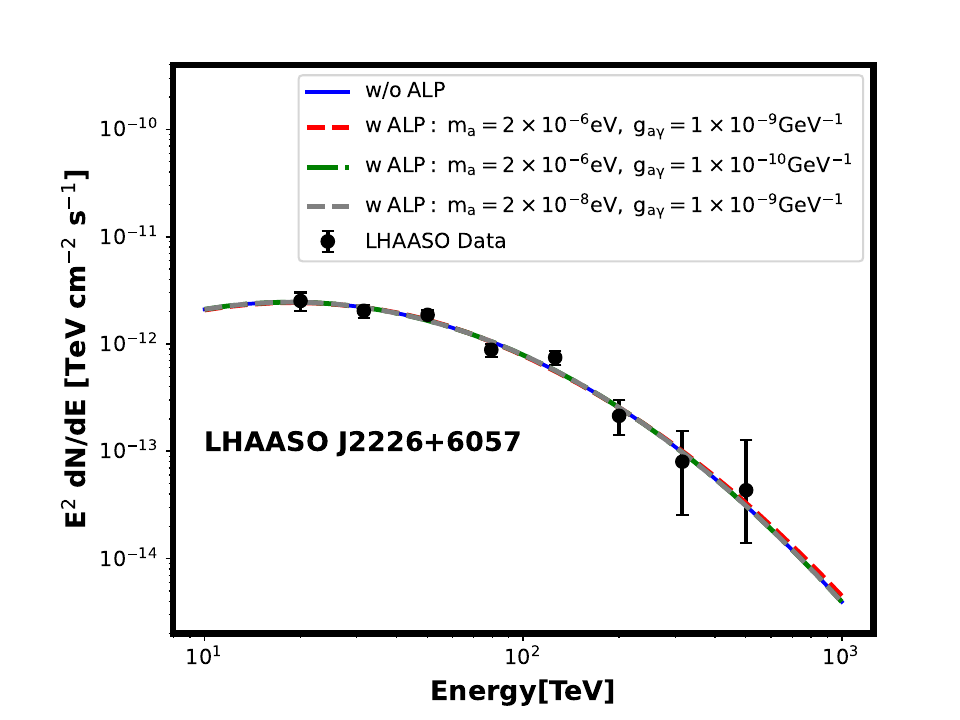}
  \caption{The best-fit spectra for the Crab Nebula, LHAASO J1825-1326, LHAASO J1908+0621, and LHAASO J2226+6057. The solid and dashed lines represent the spectra under the null and alternative hypotheses, respectively. The black points denote the photon spectra measured by LHAASO \cite{lhaaso2021peta,cao2021ultrahigh}.}
  \label{fig:Spec}
\end{figure*}

In this section, we present the constraints on the ALP parameters derived from the LHAASO observations of four Galactic sources, namely the Crab Nebula, LHAASO J2226+6057, LHAASO J1908+0621, and LHAASO J1825-1326. 
Note that each of the sources LHAASO J2226+6057, LHAASO J1908+0621, and LHAASO J1825-1326 has multiple potential candidates, as detailed in Extended Data Table 2 in Ref. \cite{cao2021ultrahigh}. The distances to these sources are important for determining the propagation length of photons within the Galactic magnetic field, thereby influencing the constraints on the ALP parameters. 
For the sake of consistency, the distances used in Ref. \cite{cao2021ultrahigh} for spectral fitting are adopted here, which are 0.8 kpc for LHAASO J2226+6057, 3.4 kpc for LHAASO J1908+0621, and 3.1 kpc for LHAASO J1825-1326.

In our analysis, we calculate the best-fit $\chi^2$ values for the four sources under the null hypothesis, yielding $\chi^2/ndf$ = 1.98, where $ndf$ denotes the number of degrees of freedom. These results indicate that the observed data aligns well with the null hypothesis without the presence of ALPs. The corresponding best-fit spectra are depicted by the blue lines in FIG. \ref{fig:Spec}. Furthermore, to illustrate the influence of ALPs on the spectrum, we have included the best-fit spectra for three ALP parameter points in FIG. \ref{fig:Spec}.

\begin{figure}[htbp]
\includegraphics[width=0.45\textwidth]{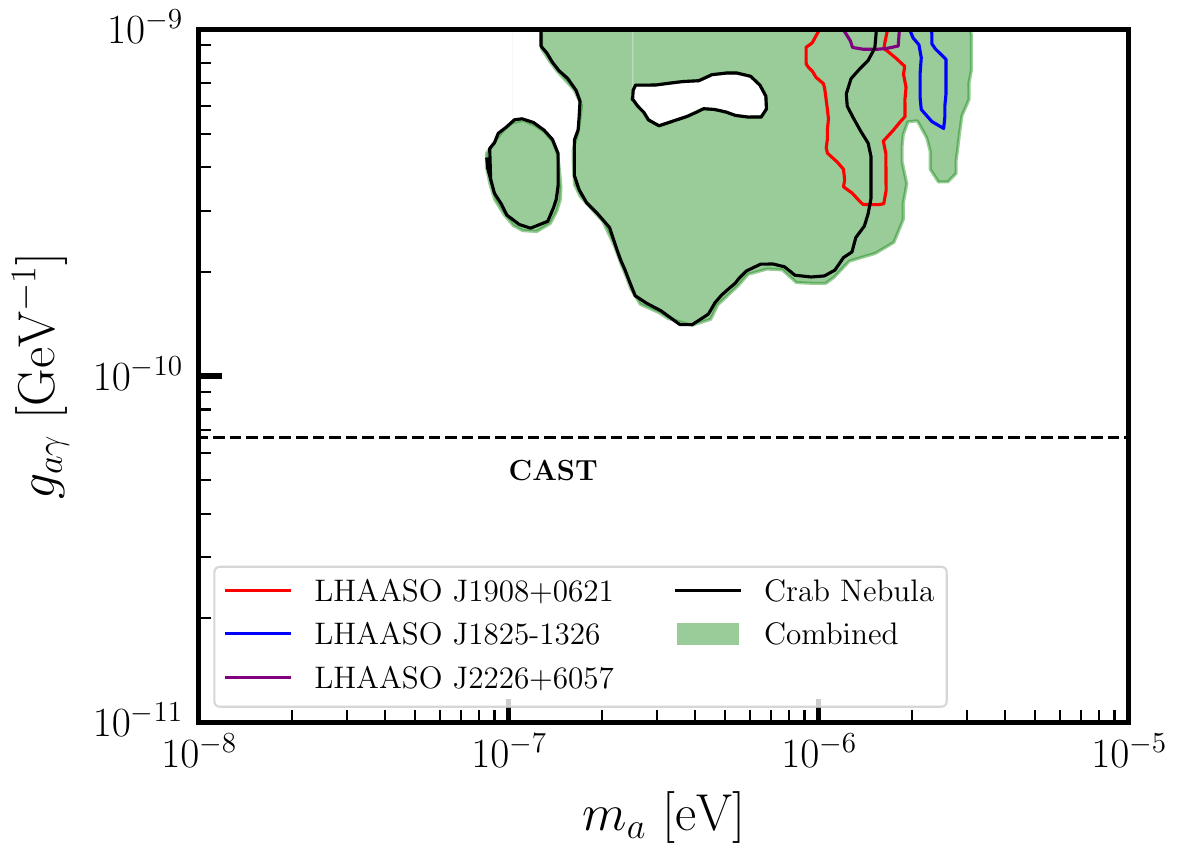}
\caption{The constraints at the 95\% C.L. in the $m_a-g_{a\gamma}$ plane resulting from the LHAASO observations of the Crab Nebula, LHAASO J2226+6057, LHAASO J1908+0621, and LHAASO J1825-1326. The black dashed line represents the constraints obtained from CAST observations, indicating $g_{a\gamma}< 6.6\times 10^{-11} \rm GeV^{-1}$ \cite{CAST:2017uph}.}
\label{resLHAASO}
\end{figure}

Using the $\rm CL_s$ method, we scan the parameter space with $m_a \in [10^{-8},10^{-5}]~\mathrm{eV}$ and $g_{a\gamma} \in [10^{-11},10^{-9}]~\mathrm{GeV}^{-1}$, and establish constraints at the 95\% C.L. for each source, as illustrated in FIG \ref{resLHAASO}. The solid lines in black, purple, red, and blue represent the constraints derived from the LHAASO observations of the Crab Nebula, LHAASO J2226+6057, 
LHAASO J1908+0621, and LHAASO J1825-1326, respectively. The most stringent constraint of $g_{a\gamma}$ is about $1.4 \times 10^{-10}$ $ {\rm GeV}^{-1}$
with the ALP mass of $m_a \sim 4\times 10^{-7} ~\mathrm{eV}$ .  

Notably, the constraints from the Crab Nebula
are considerably more stringent than those from the other sources. This can be attributed to two advantages of the LHAASO observations of the Crab Nebula.   
Firstly, the spectrum of the Crab Nebula encompasses both WCDA and KM2A results, and is precise in the energy regions of $\mathcal{O}$(TeV). In contrast, the spectra of other sources considered here only include the KM2A observations at energies above $\mathcal{O}(10)$(TeV). This indicates that it is easier to precisely determine the intrinsic spectrum of the Crab and to investigate the effects of ALP-photon oscillation in the Crab spectrum at lower energies. Secondly, the highest energy bin of the Crab spectrum reaches  $1~\mathrm{PeV}$, surpassing that of the other sources. As previously mentioned, the compensation of the ALP-photon oscillation to the absorption effect may be much more significant for higher energy photons. These characteristics enable the Crab Nebula to provide more stringent constraints.

Given that the constraints from the individual sources complement each other in the parameter space, we present the combined analysis result in  FIG. \ref{resLHAASO}.  The green region represents the combined constraint of these sources. This improves the constraints from a single source for the ALP masses above $10^{-6}~\mathrm{eV}$. 

As the observation of the Crab Nebula provides the most stringent constraints among the four sources, we conduct an analysis combing the observations from LHAASO and other experiments, including HAWC \cite{abeysekara2019measurement}, ${\rm AS\gamma}$ \cite{amenomori2019first}, HEGRA \cite{aharonian2004Crab}, MAGIC \cite{aleksic2016major}, HESS \cite{aharonian2006observations}, and VERITAS \cite{meagher2015six} for the Crab Nebula.
The $\chi^2$ for this analysis is defined as \cite{bi2021axion}
\begin{equation}
    \chi^2 =  \sum\limits _{k}\sum\limits _{i} \frac{(\widetilde{\Phi}_{k,i}-f_k^{n-1}\cdot{\Phi}_{k,i})^2}{(f_k^{n-1}\cdot \delta \Phi_{k,i})^2}+\sum \limits_{k}\frac{(f_k-1)^2}{\delta f_{k}^2},
\end{equation}
where the subscript $k$ denotes the $k$-th experimental
data, $i$ denotes the $i$-th energy bin, and $\widetilde{\Phi}$, $\Phi$ and $\delta \Phi$ are the expected value, observed value, and uncertainty of the photon flux, respectively. As the high-energy gamma-ray experiments under consideration typically have an energy resolution of approximately $\mathcal{O}(10)$ percent, the spectra measured by different experiments may not precisely match.
To derive consistent results for all experiments while accounting for this effect, we introduce additional free parameters $f$ to scale the energies of all the experiments, except for LHAASO, and add their Gaussian contributions in the $\chi^2$ function. As experimental data is often presented in the form of $E^n \frac{d N}{d E}$, $\Phi$ and $\delta \Phi$ are also scaled by a factor of $f^{n-1}$, with n=2 in this work. We take the deviations of scale factors $\delta f$ according to experimental energy resolutions, with values of 0.15 for HEGRA, MAGIC, HESS, and VERITAS, 0.14 for HAWC, and 0.12 for $\rm{AS \gamma}$. 
This approach enables us to accommodate the uncertainties arising from energy reconstruction and conduct a more comprehensive analysis incorporating data from multiple experiments.

\begin{figure}[H]
\includegraphics[width=0.45\textwidth]{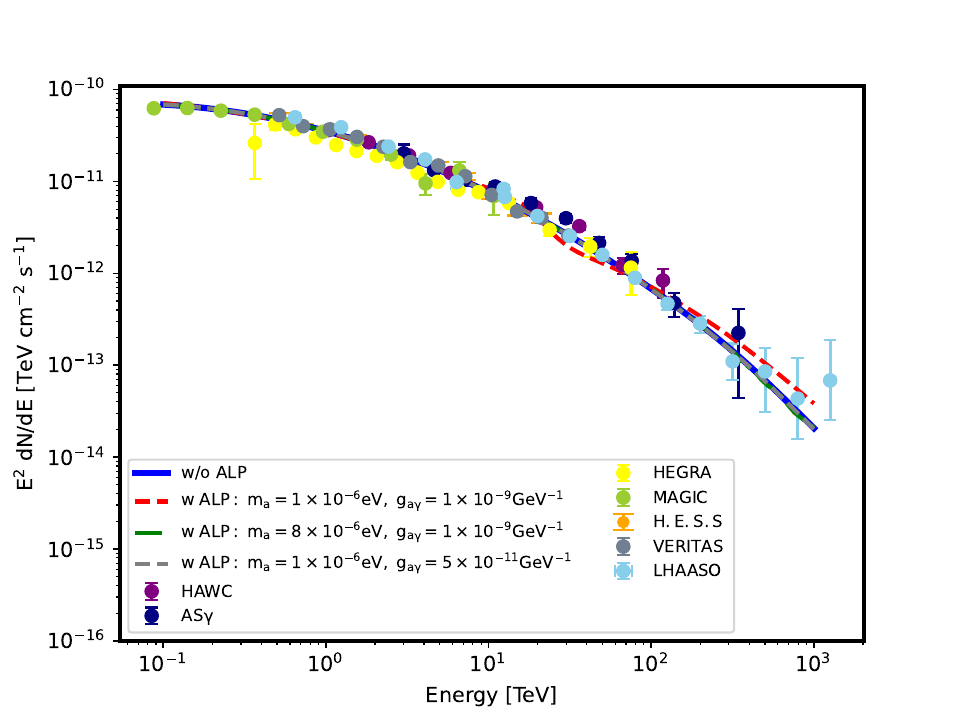}
\caption{The best-fit photon spectra for the Crab Nebula. The solid and dashed lines represent the spectra under the null and alternative hypotheses, respectively. The data points represent observations from LHAASO \cite{lhaaso2021peta}, HAWC \cite{abeysekara2019measurement}, ${\rm AS{\gamma}}$ \cite{amenomori2019first}, HEGRA \cite{aharonian2004Crab}, MAGIC \cite{aleksic2016major}, HESS \cite{aharonian2006observations}, and VERITAS \cite{meagher2015six}.}
\label{figDataCrab}
\end{figure}

\begin{figure}[H]
\includegraphics[width=0.45\textwidth]{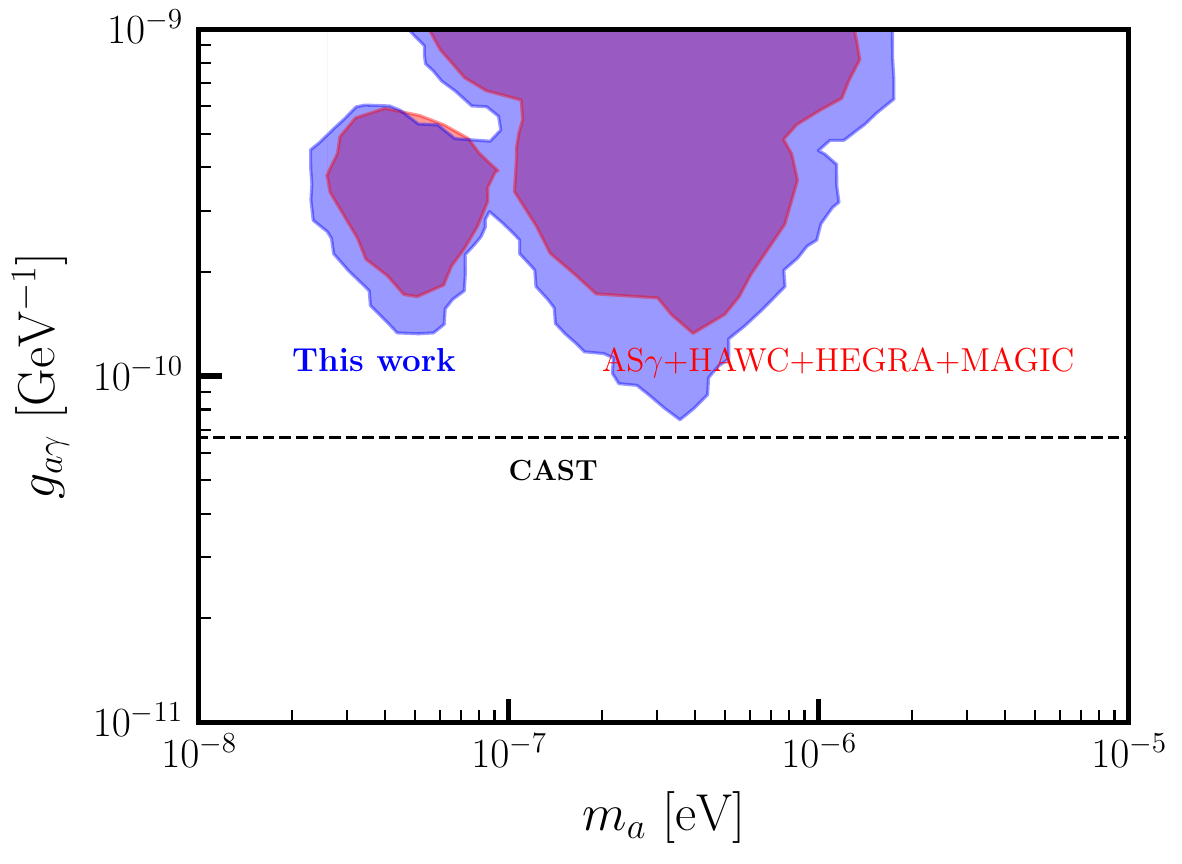}
\caption{The constraints derived from observations of the Crab Nebula, combining the data from seven experiments, including LHAASO \cite{lhaaso2021peta}, HAWC \cite{abeysekara2019measurement}, ${\rm AS{\gamma}}$ \cite{amenomori2019first}, HEGRA \cite{aharonian2004Crab}, MAGIC \cite{aleksic2016major}, HESS \cite{aharonian2006observations}, and VERITAS \cite{meagher2015six}, are depicted in the blue shaded region. For comparison, constraints from the observations of AS$\gamma$, HAWC, HEGRA, and MAGIC, as reported in Ref. \cite{bi2021axion}, are depicted in the red shaded region.}
\label{figCrab}
\end{figure}

We achieve the best-fit $\chi^2/ndf$ of 1.39 under the null hypothesis, with the corresponding spectrum illustrated in FIG. \ref{figDataCrab}. Furthermore, the best-fit spectra for three ALP parameter points and the observational data of the Crab Nebula from various experiments are depicted in FIG. \ref{figDataCrab}. Utilizing the $\rm CL_s$ method, we establish 95\% C.L. constraints on the ALP parameters, as depicted by the blue region in FIG. \ref{figCrab}. Notably, the constraints exceed the results obtained solely from the LHAASO observations. For comparison, constraints derived from observations by AS$\gamma$, HAWC, HEGRA, and MAGIC, as reported in \cite{bi2021axion}, are also included in FIG. \ref{figCrab}.  It is evident that our constraints are more stringent than those presented in Ref. \cite{bi2021axion}, and the most stringent constraint at $m_a \sim 4 \times 10^{-7}~\mathrm{eV}$ is in close proximity to the CAST constraint in \cite{CAST:2017uph}.

\section{conclusion}
\label{conclusion}
\flushend

In this paper, we study the impact of the ALP-photon oscillation effect on the gamma-ray spectra of four Galactic sources, namely the Crab Nebula, LHAASO J2226+6057, LHAASO J1908+0621, and LHAASO J1825-1326, measured by LHAASO. We consider the compensation of the ALP-photon oscillation to the absorption effect for high-energy photons, and utilize the $\rm CL_s$ method to set constraints on the ALP parameters. 

Among the four sources, the Crab Nebula provides much more stringent constraints than the other sources, due to its energy spectra spanning a wide range over three orders. By combining the data from the four sources, we find that the ALP-photon coupling larger than 
$1.4\times10^{-10}$ ${\rm GeV}^{-1}$ can be excluded for the ALP mass of $\sim 4\times10^{-7} ~\mathrm{eV}$ 
at the 95\% C.L..
Furthermore, we perform a combined analysis for the observations of the Crab Nebula from LHAASO and other experiments. Our analysis sets a limit on $g_{a\gamma}$ about $7.2\times10^{-11}$ ${\rm GeV}^{-1}$ 
for the ALP mass $\sim 4 \times10^{-7}~\mathrm{eV}$ , which is in close proximity to the CAST constraint.

\acknowledgements
Xiao-Jun Bi is supported by the National Natural Science Foundation of China under grants 12175248. Xiaoyuan Huang is supported by the National Key Research and Development Program of China (No. 2022YFF0503304), the National Natural Science Foundation of China (No. 12322302), the Project for Young Scientists in Basic Research of Chinese Academy of Sciences (No. YSBR-061), the Chinese Academy of Sciences, and the Entrepreneurship and Innovation Program of Jiangsu Province.
\balance
 \bibliographystyle{apsrev}
\bibliography{Ref}
\end{document}